# Extremely long quasi-particle spin-lifetime in superconducting aluminum using MgO tunnel spin injectors

Hyunsoo Yang[1*&], See-Hun Yang[1*], Saburo Takahashi[2,4], Sadamichi Maekawa[3,4] and Stuart S. P. Parkin[1†]

[1] *IBM Almaden Research Center, 650 Harry Road, San Jose, California 95120, USA*

[2] *Institute for Materials Research, Tohoku University, Sendai 980-8577, Japan*

[3] *Advanced Science Research Center, Japan Atomic Energy Agency, Tokai 319-1195, Japan*

[4] *CREST, Japan Science and Technology Agency (JST), Tokyo 102-0075, Japan*

[*]These authors contributed equally to this work.

[&]Present Address: Department of Electrical and Computer Engineering, National University of Singapore, 117576 Singapore

[†]E-mail: parkin@almaden.ibm.com.

**There has been an intense search in recent years for long-lived spin-polarized carriers for spintronic and quantum-computing devices. Here we report that spin polarized quasi-particles in superconducting aluminum layers have surprisingly long spin-lifetimes, nearly a million times longer than in their normal state. The lifetime is determined from the suppression of the aluminum's superconductivity resulting from the accumulation of spin polarized carriers in the aluminum layer using tunnel spin injectors. A Hanle effect, observed in the presence of small in-plane orthogonal fields, is shown to be quantitatively consistent with the presence of long-lived spin polarized quasi-particles. Our experiments show that the superconducting state can be significantly modified by small electric currents,**

**much smaller than the critical current, which is potentially useful for devices involving superconducting qubits.**

The interplay of magnetism and superconductivity has long been of great interest[1-3]. The influence of magnetism on superconductivity has been studied by the proximity effect, i.e. the placement of magnetic materials adjacent to superconducting materials, on various lengths scales from the macroscopic, where magnetostatic fields are important, to the microscopic, where exchange effects become relevant, strongly suppressing superconductivity[4-8]. These exchange effects are mitigated by the insertion of ultra-thin tunnel barriers between the ferromagnet and superconductor[9-11], allowing spin-polarized carriers to be injected into the superconductor. By adding a second such insulator-ferromagnet combination on the far side of the superconductor, spin polarized carriers can be accumulated in the superconductor. The accumulation is predicted to take place only when the ferromagnetic layer moments are antiparallel to each other, and, if large enough, to lead to a suppression of the superconducting energy gap[12]. Here we show that in large area planar double magnetic tunnel junctions formed with MgO tunnel barriers, a highly efficient spin injector[13,14], the superconducting energy gap is indeed reduced in aluminum for the *antiparallel* configuration. This leads to an oscillatory variation of the tunneling magnetoresistance with bias voltage for temperatures well below the superconducting transition temperature of the Al layer. Modeling of the experimental results shows that these results can only be accounted for by a dramatic enhancement of the quasi-particle spin lifetime $\tau_S$ in the superconducting state *by several orders of magnitude* as compared to the normal state.

The spin relaxation time in the normal state of Al ($\tau_N$) has been intensively studied with measured values ranging from 0.1 to 10 nsec depending on the film thickness and quality[15-18]. However, the experimental situation for the superconducting (SC) state is less clear with one group inferring a shorter spin lifetime in the SC state[19] than in the normal state and others assuming that the spin lifetime is unchanged[20,21]. Theoretically, it is anticipated that $\tau_S$ should be increased because the group velocity of quasi-particles injected into the bottom of the quasi-particle conduction band just above the superconducting energy gap, is small, thereby leading to reduced interaction probabilities. This can be restated more clearly as being a result of the spin-bottleneck to quasi-particle relaxation due to spin-charge separation[22,23] at the energy gap edge, where the quasi-particles have only spin and no charge, leading to reduced spin-orbit scattering and therefore an enhanced spin relaxation time[24].

Experiments on the interplay between spin accumulation and superconductivity have been carried out by two groups using single electron transistor (SET) devices consisting of small Al islands electrically connected on their top surface by closely and laterally spaced ferromagnetic Co nanowire electrodes[20,25,26]. However, the SET experimental configuration has led to controversy over their interpretation due to magnetic fringing fields from the Co[21,26,27]. The effect of fringing fields on the superconducting properties is especially important for thicker SC Al layers[28]. Here we use a vertical geometry and planar tunnel junctions with large areas ~700 × 700 $\mu m^2$ so that fringing fields from the $Co_{70}Fe_{30}$ ferromagnetic (F) electrodes are negligibly small. Large areas are also used so that the tunnel barriers, formed from MgO, could be as thick as possible to ensure the absence of any exchange proximity effect on the superconducting Al layer from the F electrodes. The double tunnel junction (DTJ) devices were fabricated by dc magnetron sputter deposition using a series of four *in-situ* shadow masks to form a junction, as schematically shown in Fig. 1c. The direction of the magnetization of the lower F electrode was fixed using exchange bias from an IrMn antiferromagnetic layer on which the F layer was deposited. The exchange bias field was ~1500 Oe for the temperature range of interest (~ 0.25–2.5 K). Thus, the moment of the upper F electrode could be independently oriented parallel (P) or antiparallel (AP) to the lower electrode's moment in small magnetic fields (~ ±300 Oe) to create well defined magnetic configurations of the DTJ. These magnetic fields also serve to negate the effect of any small perturbing orthogonal in-plane magnetic fields which might arise from, for example, inhomogeneities in the magnetization of the electrodes. Such fields could otherwise depolarize the accumulated spins especially for very long spin lifetimes (see Methods).

Conductance measurements were carried out in a $^3$He refrigerator using standard ac lock-in four-probe techniques for a series of samples in which the MgO and Al layer thicknesses were varied (see Methods). Typical bias voltage (*V*) dependences of the conductance in the P and AP states ($G_P$ and $G_{AP}$, respectively) are shown in Fig. 1a for a DTJ with an MgO thickness of ~3.3 nm and an Al layer thickness of 4.5 nm. Data are shown for several temperatures varying from 0.25 K to 2.5 K. The conductance data at 0.25 K show evidence for a high quality superconducting tunnel junction with a well defined superconducting energy gap. A distinctive feature of these data is that the superconducting energy gap is slightly smaller for the AP as compared to the P magnetic configuration. The difference between the SC energy gaps in the AP and P states ($\Delta_{AP}$ and $\Delta_P$, respectively) decreases with increasing temperature. The data also show a reduced quasi-particle peak intensity near $V/2 \sim \pm \Delta_{AP}/e$ for the AP state. From the difference in the conductance curves we can define a tunneling magnetoresistance, TMR = $(G_P - G_{AP})/G_{AP}$, which is plotted in Fig. 1b as a function of bias voltage at various temperatures. The TMR data at 0.25 K exhibit a distinctive oscillatory dependence on the bias voltage. In particular, as the bias voltage is increased from zero, for either positive or negative V, a large negative TMR is observed with a peak value ~ −55% at ~ ±0.35mV. The TMR changes sign and becomes positive (~ +23% at ~ ±0.58mV) as V/2 approaches $\Delta_P/e$, reaching zero at slightly higher V. As the temperature is increased the TMR decreases to zero as the temperature approaches the superconducting transition temperature of the Al layer ($T_c$ ~ 2.3 K), which shows that the TMR originates from the superconducting state of the Al layer. Similar results were obtained in several devices in which the MgO thickness was varied, as discussed later.

A ferromagnet in direct contact with a superconducting layer causes suppression of the superconductivity through the exchange interaction at the interface[4,6,7]. When superconducting layers, thin compared to their corresponding coherence length, are sandwiched between F layers, it has been found that $T_c$ (as inferred from resistance versus temperature measurements) is suppressed for both P and AP configuration of the F layers but that the suppression is slightly greater for the P configuration[4,6,7]. This is contrary to our results. Indeed, by inserting a leak-proof thick tunneling barrier between the F and SC layers, as in our experiments, the proximity effect can be ruled out[1].

Our results are consistent with a model in which non-equilibrium spin density is accumulated in the SC layer when spin-polarized current is injected through tunnel barriers from F electrodes in the AP configuration[12]. As illustrated in Fig. 2 no spin-density $S_P$ is accumulated in the P configuration but spin density $S_{AP}$ is formed in the SC in the AP state when the spin-relaxation time of the spin-polarized quasi-particles is sufficiently long. This is because the tunneling is spin-dependent: for CoFe/MgO/Al the tunneling current is strongly majority spin-polarized so that for the symmetric DTJs used here, electrons can readily tunnel into and out of the SC Al layer in the P but not in the AP state (see Fig. 2). The magnitude of $S_{AP}$ clearly depends sensitively on the spin relaxation time $\tau_S$ for a given spin injection rate. The lower the spin injection rate the longer the spin relaxation time needed to establish significant $S_{AP}$. We will show in the following that the magnitude of $\tau_S$ must be extremely large to account for our data.

The accumulated spin density $S$ in the SC arises from an imbalance between the populations of the up-spin and down-spin quasi-particles, which corresponds to a small shift in their chemical potentials by $\pm\delta\mu$ from their equilibrium values (see Fig. 2). In the presence of spin-orbit scattering of the quasiparticles by impurities or grain boundaries in the SC layer, the accumulated spin density $S$ is given by (see Supplementary Information for details),

$$S \approx (1-b/2)\int_\Delta^\infty \mathcal{D}_S(E)[f(E-\delta\mu)-f(E+\delta\mu)]dE + (b/2)\delta\mu \qquad (1)$$

where $\mathcal{D}_S(E)$ is the density of states with the superconducting energy gap $\Delta$, $b = \hbar/(3\tau_N \Delta_0)$ is the spin-orbit parameter, $\Delta_0$ is the BCS gap at zero temperature, and $\tau_N$ is the spin relaxation time in the normal state. The use of the Fermi distribution function is a good approximation because the energy relaxation time (<1 μsec)[29] is significantly shorter than the spin relaxation time in the Al SC.

On the other hand, the spin-up and spin-down tunneling currents between each F

electrode and the SC layer, which flow by application of the bias voltage $V$, are proportional to the spin-dependent density of states in the respective layers multiplied by the corresponding tunneling matrix elements[30]. The spin injection rate $(dS/dt)_{inj}$ due to the spin-polarized tunneling current is balanced by the spin relaxation rate $(S/\tau_S)$ due to spin-flip scattering of the injected quasi-particles by spin-orbit interactions in the SC layer. Thus, the spin accumulations for the P and AP configurations, denoted by $S_P$ and $S_{AP}$, respectively, are given by,

$$S_P = 0, \qquad S_{AP} = P' \int_{\Delta_{AP}}^{\infty} \mathcal{D}_S(E)\left[f(E-eV/2) - f(E+eV/2)\right] dE \qquad (2)$$

where $P'$ is the effective spin polarization of the tunneling current through the double tunnel junction, $\Delta_{AP}$ is the superconducting energy gap in the AP state, and $f(E \pm eV/2)$ are the Fermi distribution functions of the left (−) and right (+) ferromagnetic electrodes[12]. In the absence of spin relaxation in the SC layer, $P'$ is identical to the spin polarization of the tunneling current $P$ measured, for example, in F-I(insulator)-SC junctions using the superconducting tunneling spectroscopy (STS) technique. In the presence of spin relaxation, the effective spin polarization of the tunneling current is reduced and is given by the relation $P' = P/(1+\Gamma_S)$, where $\Gamma_S = (\gamma_t \tau_S)^{-1}$ is a spin relaxation parameter which itself depends on both the spin relaxation time $\tau_S$ and the quasi-particle injection rate $\gamma_t$ [12,31] (see Supplementary Information, section 1, for details).

The spin accumulation in the SC layer results in a reduction of the SC energy gap. The effect of $S_{AP}$ on $\Delta_{AP}$ can be described using a modified gap equation in which the effect of the chemical potential shifts $\pm\delta\mu$ is incorporated in the BCS gap equation[12] (see Supplementary Information, section 2, for details). By solving self-consistently this gap equation together with the results of equating (1) and (2), which allow us to find a relationship for $\Delta$ in terms of $\delta\mu$ and $V$, we finally obtain $\Delta_{AP}$ and $\delta\mu$ as functions of the bias voltage.

We use $\Delta_{AP}$ and $\delta\mu$ to calculate the tunneling currents $I_P$ and $I_{AP}$ for the parallel and antiparallel states. The derivatives of these currents with respect to $V$ give the tunneling conductances $G_P$ and $G_{AP}$, and, thereby, the voltage dependence of TMR (see Supplementary Information section 2 for details). Fits to the TMR data shown in Fig. 1d require very large values of $\tau_S$ (~ 0.1 msec) for $|V/2| < \Delta_0$ which are five to six orders of magnitude larger than $\tau_N$, where we obtain the latter from STS experiments on related F-I-SC junctions. To obtain a significant TMR oscillation from spin imbalance, $1/(\gamma_t \tau_S) \sim < 1$ is required, that is, a high tunneling injection rate or a long spin relaxation time. The detailed dependence of $\tau_S$ on the applied voltage is shown in Fig. 3b for various spin injection rates as defined by the normal state spin relaxation parameter of $\Gamma_N$. $\Gamma_N$ is given by $R_T A d_S / (2\rho_N \lambda_N^2)$ where $R_T A$ is the product of the tunnel barrier resistance $R_T$ and area $A$, $d_S$ is the SC thickness, $\rho_N$ is the resistivity, and $\lambda_N = \sqrt{D_N \tau_N}$ $(=\lambda_S)$ is the spin diffusion length of Al, where $D_N$ is the diffusion constant in the normal state. Since the value of $\Gamma_N$ is ~ $10^6$ in our planar tunnel junctions due to the very low tunneling rates through the thick MgO barriers, this means that we cannot observe a significant TMR unless $\Gamma_S = (\tau_N/\tau_S) \Gamma_N$ becomes of the order of or is smaller than unity, i.e., $\tau_S$ is about $10^6$ times longer than $\tau_N$. The large value of $\Gamma_N$ is consistent with vanishing TMR above $T_c$, as shown in Fig. 1b, i.e. $P'$ becomes negligibly small due to the large $\Gamma_N$ and, therefore, there is no spin accumulation in the normal-state.

We can justify the very high values of $\tau_S$ of ~ 0.1 msec for $|V/2|<\Delta_0$ from the following considerations. The electron spin relaxation is dominated by spin-orbit scattering at low temperatures in both the normal and the superconducting states, whereas phonon scattering is only important in the normal state far above $T_c$ [32,33]. The spin relaxation time in the superconducting state can therefore be written as,

$$\tau_S(T) = \frac{\chi_S(T)/\chi_N}{2f(\Delta)} \tau_N \qquad (3)$$

where $\chi_S/\chi_N$ is the spin susceptibility of the superconductor normalized to that of the normal-state, and $f(\Delta)$ is the Fermi distribution function at the energy gap $\Delta$ (see Supplementary Information for details). At low enough temperatures ($T \ll T_c$), $\chi_S/\chi_N$ remains finite due to the spin-orbit scattering of quasi-particles in the Al layer and is given by $b/2$ [34], where $b$ is measured in STS studies of closely related CoFe/MgO/Al tunnel junctions[35]. We interpolate the susceptibility between $T=0$ and $T=T_c$ by using the well known Yosida function[36], $\chi_S^0/\chi_N$. Thus, we conclude that $\chi_S/\chi_N$ can be written as $(1-b/2)\chi_S^0/\chi_N + b/2$, since $\chi_S^0/\chi_N \rightarrow 0$ as $T \rightarrow 0$. Thus, this interpolated formula accounts for both the finite spin-susceptibility at low temperatures and the normal state result. Since previous theories[31] did not take into account spin-orbit effects on $S$ and $\chi_S/\chi_N$, they gave only a modest increase of $\tau_S$ compared to $\tau_N$. The inclusion of the spin-orbit term in $S$ and $\chi_S/\chi_N$ is critical for explaining the extremely long spin lifetimes at low temperatures that we find in our planar tunnel junction devices.

In Fig. 3e, we plot the temperature dependence of $\tau_S$ calculated in this way for various values of $b$ (we measure $b \sim 0.02$ in the samples used here). The very strong temperature dependence of $\tau_S$ is associated with the development of the superconducting energy gap $\Delta$. As the temperature decreases below $T_c$ the number of excited quasi-particles above $\Delta$ decreases and quasi-particles are populated only near the gap edge, where they have zero charge, spin one-half, and very small velocity. A slowly moving quasi-particle takes a longer time to be scattered by impurities compared with an electron in the normal state, so that the momentum-scattering time as well as the spin-flip scattering time become longer in the superconducting state. In addition, in the presence of spin-orbit scattering the spin susceptibility is finite, while the superconducting energy gap is unaffected (known as the Anderson theorem[37]) and $f(\Delta)$ vanishes at low temperatures. The spin-orbit scattering decreases the spin life-time in the normal state. In the superconducting state the Cooper pair condensate is comprised of spin singlets, i.e. spin zero, in the absence of spin-orbit scattering. When the spin-orbit interaction is turned on, the spin is no longer a good quantum number[34]. In the presence of a small external perturbation, such as electrical spin injection (or a magnetic field), some Cooper pairs become virtually excited so as to create quasi-particles. These have a finite spin[38], which gives rise to a finite spin susceptibility and to spin accumulation at low temperatures. As a result, the spin lifetime becomes extremely long in the superconducting state.

In Fig. 3a-d, the calculated values of $S_{AP}/\Delta_0$, $\tau_S$, $\Delta/\Delta_0$ and $P'$ are plotted as functions of bias voltage for various values of $\Gamma_N$. Here we take $\Delta_0 = 0.27$ meV and $\tau_N = 41$ psec ($\tau_N = \hbar/(3\Delta_0 b)$)[17]. The gap $\Delta_P$ in the P configuration has a BCS gap value $\Delta_0$ since there is no spin accumulation. In the AP configuration a pronounced bias dependence of these quantities is shown; a spin accumulation $S_{AP}$ appears when $|V/2|>\Delta_0$ in which the gap $\Delta_{AP}$ and the spin lifetime $\tau_S$ are suppressed. As the value of $\Gamma_N$ is increased i.e. the tunneling rate is decreased, the magnitude of the spin accumulation is reduced, so that $\Delta_{AP}$ approaches $\Delta_P$ and $\tau_S$ decreases towards $\tau_N$. The reduction of TMR with increasing $\Gamma_N$ is evident in the experimental results shown in Fig. 4. The effective spin polarization $P'$ is a measure of the spin accumulation in the Al SC. As $\Gamma_N$ is increased, $P'$ is decreased for $|V/2|>\Delta_0$ and $S_{AP}$ is correspondingly decreased. For a given value of $\Gamma_N$, the application of bias voltage decreases $P'$ for $|V/2|>\Delta_0$. This is due to the fact that the spins accumulated when $|V/2|>\Delta_0$ suppress the superconductivity of Al which thereby reduces the spin lifetime. Consequently, a smaller proportion of the injected spins now contribute to the spin accumulation. This accounts for the saturation of the spin accumulation $S_{AP}$ with increasing bias voltage for $|V/2|>\Delta_0$ as seen in Fig. 3a.

The measured dependence of the TMR in DTJs in which the MgO thickness was varied from 3.1 to 3.7 nm is shown in Fig. 4. The TMR decreases with increasing MgO thickness as expected from the above model. For a given voltage, as the MgO thickness and the corresponding $R_T A$ value increase, the spin injection rate and therefore the spin accumulation is decreased. Values of TMR calculated using measured $R_T A$ values but otherwise the same parameters as for the data in Fig. 1d are in good agreement with the experimental data.

Finally, we use the Hanle effect to confirm the long spin lifetimes of the accumulated spin polarized quasi-particles in the Al superconducting layer. To carry out this experiment, an in-plane field $H_\perp$, perpendicular to the direction of injected spins, is applied for the AP configuration of the magnetic electrodes. Note that a perpendicular out-of-plane field that is typically used in Hanle studies[39,40] significantly suppresses superconductivity due to the small critical field of Al. When $H_\perp$ is small enough not to upset magnetization of electrodes, the injected spins precess around $H_\perp$ within the Al SC, and the spins become depolarized. The longer the spin-lifetime, the smaller is $H_\perp$ that suppresses the spin accumulation, since a greater number of coherent spins contribute to the Hanle dephasing[39,40]. When a magnetic field $H_\parallel$ is applied in a direction collinear with that of the accumulated spins in addition to $H_\perp$, the accumulated spins will precess about the direction of the total applied field (what is called the oblique Hanle effect)[41] (see Methods). The suppression of the spin accumulation results in the restoration of the SC gap. In our experiments $H_\parallel$ is varied for a fixed value of $H_\perp$. Since the spin accumulation varies as $H_\parallel^2/(H_\parallel^2 + H_\perp^2)$ [41], $\Delta_{AP}$ therefore should exhibit a distinctive peak near $H_\parallel = 0$ (see Fig. 5b and Supplementary Information). As shown in Fig. 5, we find clear evidence for such a Hanle dephasing peak in the presence of small perpendicular fields (no peak is observed when $H_\perp \approx 0$). In our experiment, the applied in-plane fields $H_1$ and $H_2$ are not exactly aligned along $H_\parallel$ and $H_\perp$ but deviate from these directions by the angles $\theta_\parallel$ and $\theta_\perp$, respectively (see inset in Fig. 5a). This deviation results in an asymmetric switching of the spin accumulation (i.e. an abrupt change for P→AP but a gradual change for AP→P) and an asymmetry in the Hanle dephasing peak (see Fig. 5a,b). These asymmetries are well reproduced by our model (see Fig. 5c,d and Supplementary Information). A detailed analysis of our experimental results shows very good agreement with the long spin-lifetimes inferred from our TMR data.

The extremely long spin lifetimes of quasi-particles in the superconducting state of Al which we have determined suggests that these could be useful for a wide range of applications in spin-based computing devices and superconducting materials. For example, the information (or the entanglement) encoded in long lived spins[42] can be more effectively preserved and manipulated if the superconducting Al is used in quantum computing devices. In addition, the double magnetic tunnel junction we studied may be potentially used in spintronic devices such as a very low power switch or a memory device since the application of a tiny voltage can control the conductance of the device.

**Methods**

Planar DTJs of the form F1/ MgO/ SC/ MgO/ F2 were fabricated on Si(100)/25nm $SiO_2$ substrates using a sequence of metal shadow masks in an automated high vacuum deposition system without breaking vacuum. A highly textured MgO(100) tunnel barrier is formed by first depositing a thin 0.8 nm thick Mg layer on the lower exchanged-biased magnetic electrode F1, which is formed from 10 nm MgO/10 nm Ta/ 25 nm $Ir_{24}Mn_{76}$/ 3.5 nm $Co_{70}Fe_{30}$ [13]. The Mg layer is used to prevent oxidation of the underlying $Co_{70}Fe_{30}$ layer. The MgO barrier is then formed by reactive sputtering of MgO from a Mg sputter target in an $Ar-O_2$ environment. After growing the Al SC layer (~4.5 nm thick), the second MgO tunnel barrier (~2.7 nm) was formed by first depositing a thin metallic Mg layer (~0.8 nm) and then by reactive sputter deposition of MgO. Finally, the upper magnetic electrode F2 was formed from 15 nm $Co_{70}Fe_{30}$ followed by a 5 nm Ru capping layer. The Al superconducting layer was doped with ~5 atomic % silicon to increase $T_c$.

The spin accumulation is very sensitive to magnetic fields that are orthogonal to the magnetization direction due to Hanle precession when the spin relaxation time is very long but only when the in-plane magnetic field is close to zero. If there is an in-plane magnetic field, which is large compared to any orthogonal fields, then the spin will precess about the vectorial sum of these fields, so effectively preserving the sign of the spin accumulation (although with a reduced magnitude). In our experiments we use external magnetic fields of either +300 Oe and −300 Oe applied along the easy axis magnetization direction to maintain the P and the AP configurations, respectively. These fields are much larger than any likely perpendicular magnetic fields. However, the collinear magnetic fields which we use are too small to appreciably affect the superconducting energy gap of the Al layer.  Moreover, since the damping parameter is small in superconducting Al ($\alpha \sim 1/T_1$ where $T_1$ is the longitudinal spin life time), then precession of the accumulated spins caused by any small perpendicular fields will continue about the collinear field direction but without spin relaxation.

In the oblique Hanle effect the spin accumulation decreases slowly with respect to $H_\perp/H_{//}$, where $H_{//}$ and $H_\perp$ are the collinear and perpendicular fields to the magnetization direction, respectively.  In particular, the decrease of spin accumulation varies as $\sim H_\perp^2/(H_\parallel^2 + H_\perp^2)$ when the spin lifetime is very long[41].  For $H_{//}$= 300 Oe, a perpendicular field equivalent to the earth's magnetic field (~ 0.4 Oe) would decrease the spin accumulation by a negligible amount of only ~0.00018 %.  Even if there is a perpendicular magnetic field of as much as 17.6 % of the applied collinear field this would decrease the spin accumulation by only ~3 %.

**Acknowledgements**

We would like to thank B. J. van Wees for useful discussions. This work is partially supported by DMEA, Grant-in-Aid for Scientific Research from MEXT and the Next Generation Supercomputer Project, Nanoscience program, MEXT, Japan.

**Author Contributions**

H.Y. and S.S.P.P. initiated this work. H.Y. carried out the electrical transport experiments. S.-H.Y. grew the samples, and performed the Hanle effect experiment and the numerical calculations. S.T. and S.M. developed the theoretical models. S.S.P.P. supervised and led this research project. All authors wrote, edited the paper and discussed the data and the results.

**Additional Information**

Supplementary Information accompanies this paper on www.nature.com/naturematerials. Reprints and permissions information is available online at http://npg.nature.com/reprintsandpermissions. Correspondence and requests for materials should be addressed to sspp. (parkin@almaden.ibm.com).

**Figure 1.** Experimental conductance and TMR data and comparison with model. Normalized conductance (**a**) and TMR curves (**b**) as a function of bias voltage for various temperatures. The structure is composed of CoFe/3.3 nm MgO/4.5 nm Al/3.3 nm MgO/ CoFe for which $\Gamma_N$ is calculated to be ~$2\times10^6$ using values of $R_T A$~$10^8$ $\Omega\mu m^2$, $\rho_N$~10 $\mu\Omega$cm and $\lambda_N$~1 $\mu$m [43,44]. Blue (red) line corresponds to the conductance in the P (AP) configuration. $T_c$ of the Al layer is ~2.3 K. For clarity, the data at each temperature are offset from each other by 0.5 in (**a**) and 25% in (**b**). **c,** Illustration of a planar double tunnel junction structure. **d,** Comparison of calculated TMR versus bias voltage (solid lines) with the experimental data (open circles) at 0.25 K. Magenta line corresponds to a calculation without consideration of depairing[13] ($\zeta_{P(AP)}=0$) or quasi-particle lifetime broadening[28] ($\Gamma^{LT}_{P(AP)} = 0$) effects. Dark blue line includes these effects with corresponding fitting parameters for the P(AP) states shown in the figure.

**Figure 2.** Schematic diagram of superconducting gap suppression and spin accumulation in a double tunnel junction composed of two ferromagnet electrodes (F1 and F2) and a superconducting middle electrode (SC). Energy dependence of the spin polarized density of states in F1, F2 and SC when the SC layer is its superconducting state. Blue and red correspond to majority and minority spin polarized density of states, respectively. The P and AP configurations of F1 and F2 are shown in the left and right columns, respectively. The dashed lines represent the electrochemical potential (ECP): Blue and red dashed lines correspond to the ECPs of the up and down spins in SC. When a voltage $V$ is applied between F1 and F2, spin-polarized tunnel current flows across the junctions. In the P configuration, the up (down) spin currents at the left and right junctions are balanced with each other, and the ECPs of the up and down spins has no shift, so that there is no suppression of the energy gap.   In the AP configuration the up (down) spin currents at the junctions are imbalanced to yield the ECP shift by $\pm\delta\mu$ as indicated the blue and red dashed lines, by which spin accumulation takes place as shown by quasi-particles electrons with up spins and holes with down spins, so that the gap is suppressed, i.e., $\Delta_{AP}<\Delta_P$.   **a,** When $eV/2 = \Delta_{AP}$, the tunnel current in the P configuration is smaller than in the AP configuration because of $\Delta_P>eV/2$, so that $G_P<G_{AP}$ and the TMR is negative.   **b,** When $eV/2 = \Delta_P$, the tunnel current in the P configuration is larger than in the AP configuration because of $eV/2>\Delta_{AP}$, so that $G_P>G_{AP}$ and the TMR is positive.

**Figure 3.**   Modeling of experimental data to determine spin lifetime in the superconducting Al layer.   The bias voltage dependence in the AP configuration of the spin accumulation $S_{AP}/\Delta_0$ (**a**), the spin relaxation time $\tau_S$ (**b**), the effective spin polarization $P'$ (**c**), and the superconducting energy gap $\Delta/\Delta_0$ (**d**), at 0.25 K for various $\Gamma_N$.   Also shown in (**d**) is $\Delta/\Delta_0$ in the P state (dashed line).   All parameters are the same as those used in the calculation of Fig. 1d except for $\Gamma_N$ which is varied.   $\Gamma_N = 2\times10^6$ corresponds to the calculation in Fig. 1d.   **e**, Calculated spin lifetime in Al versus temperature normalized to the superconducting transition temperature $T_c$ for several values of $b$, the spin-orbit interaction parameter.   $b = 0.02$ (blue line) is used in the calculation in Fig. 1d and Fig. 4.

**Figure 4.**   Tunnel barrier thickness dependence of TMR.   Measured TMR values shown as open symbols – circles and squares correspond, respectively, to the maximum and minimum values (peaks and dips in Fig. 1b) – as a function of MgO thickness.

Corresponding calculated values are shown as solid symbols.

**Figure 5.** Observation of the Hanle effect in F-I-SC-I-F. **a,b**, Measurement of the suppression of the superconducting energy gap by spin accumulation at 0.25 K. The voltage separating the two conductance peaks in the conductance versus voltage curves (e.g. Fig. 1a) is plotted as a function of $H_1$ for $H_2 = 0$ and 12 Oe. Black and red curves correspond, respectively, to $H_1$ varied from + to − fields and vice versa. $H_1$ and $H_2$ correspond to in-plane fields nominally aligned parallel and orthogonal to $\mathbf{H}_K^{free}$, respectively. The inset in (**a**) describes the directions of free layer magnetization $\mathbf{M}_{free}$, in-plane applied fields $\mathbf{H}_1$ and $\mathbf{H}_2$, magnetic easy axis $\mathbf{H}_K^{free}$, and the fields $\mathbf{H}_\parallel$ and $\mathbf{H}_\perp$ that are collinear and perpendicular to $\mathbf{H}_K^{free}$. $H_\perp$ and $H_\parallel$ are related to $H_1$ and $H_2$ by $H_\parallel = H_1 \cos\theta_\parallel + H_2 \sin\theta_\perp$ and $H_\perp = H_1 \sin\theta_\parallel + H_2 \cos\theta_\perp$ where $\theta_\parallel$ ($\theta_\perp$) is the angle between $\mathbf{H}_\parallel$ and $\mathbf{H}_1$ ($\mathbf{H}_\perp$ and $\mathbf{H}_2$). Prior to each measurement, a reset field, $H_1 = 5000$ Oe, is applied. Modeling of these data are shown for $H_2 = 0$ in (**c**) and 12 Oe in (**d**) with the corresponding calculated normalized and projected (on $\mathbf{H}_K^{free}$) free layer magnetization $M_{free}^\parallel / M_{free} (= \cos\theta_{free})$ versus $H_1$ curves shown as insets where $\theta_{free}$ is the angle between $\mathbf{M}_{free}$ and $\mathbf{H}_K^{free}$ (see Section 3 at Supplementary Information).

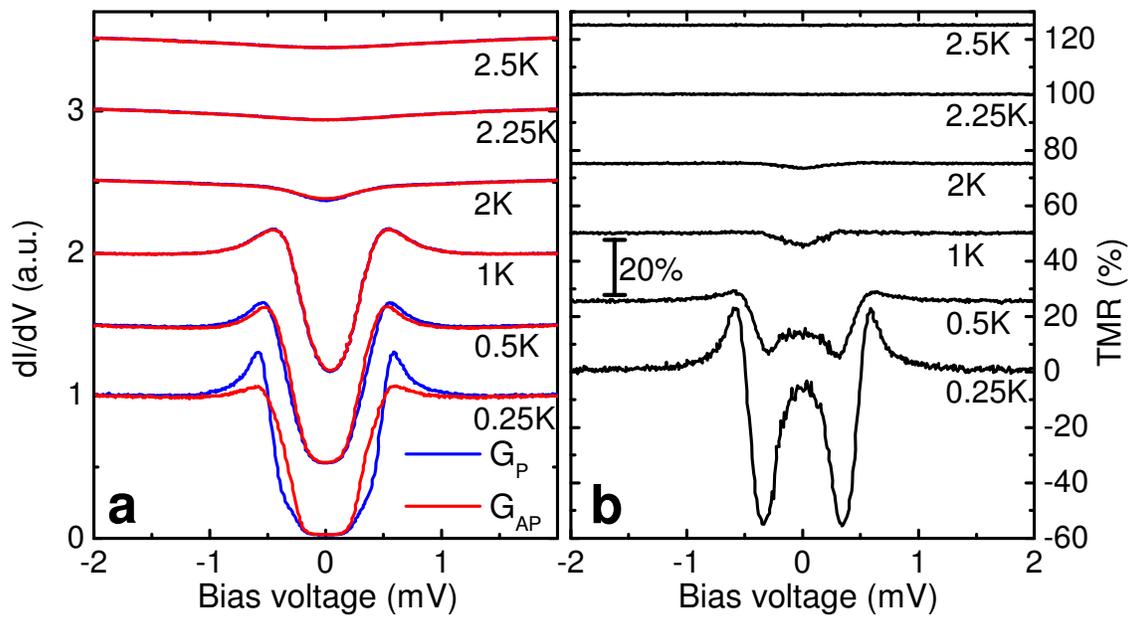
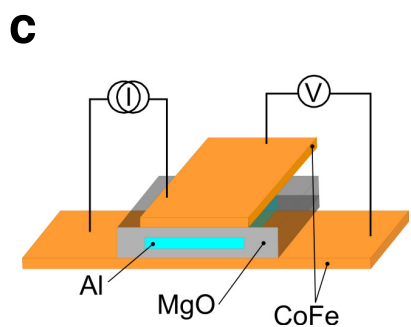
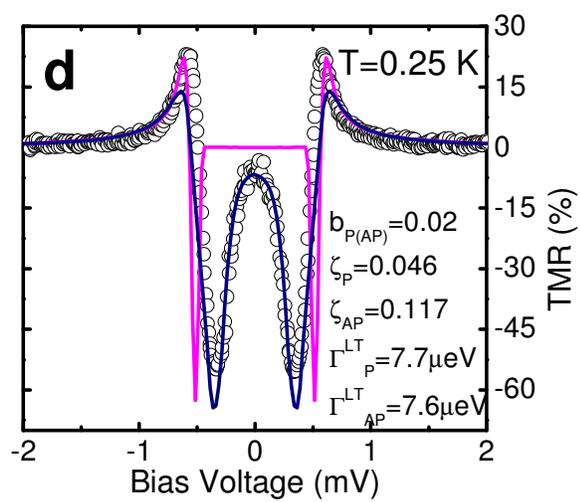

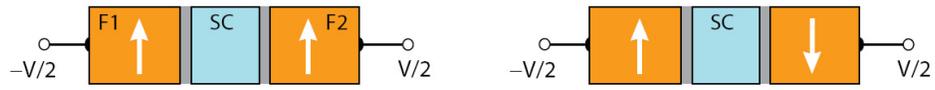

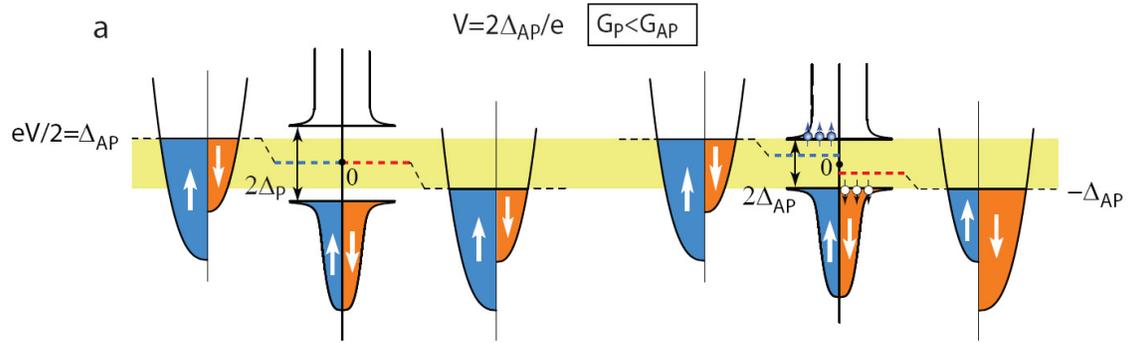

a  $V = 2\Delta_{AP}/e$  $G_P < G_{AP}$

$eV/2 = \Delta_{AP}$

$2\Delta_P$

$-\Delta_{AP}$

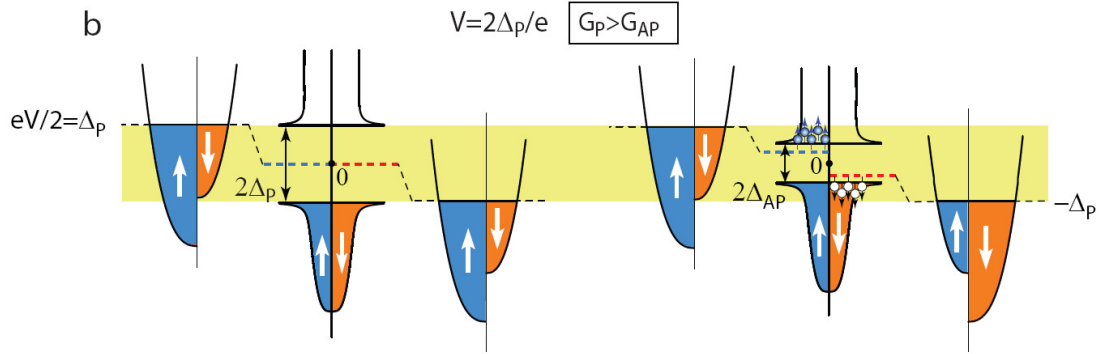

b  $V = 2\Delta_P/e$  $G_P > G_{AP}$

$eV/2 = \Delta_P$

$2\Delta_P$

$2\Delta_{AP}$

$-\Delta_P$

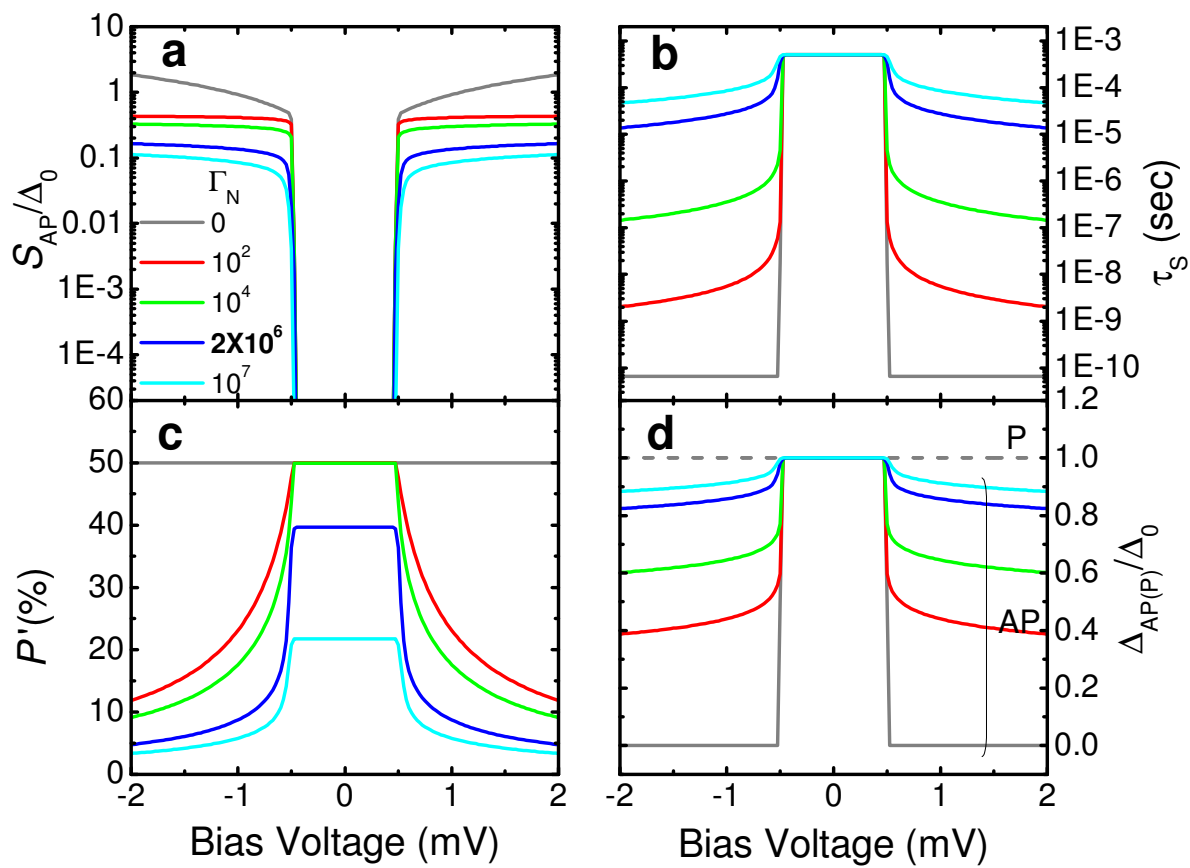
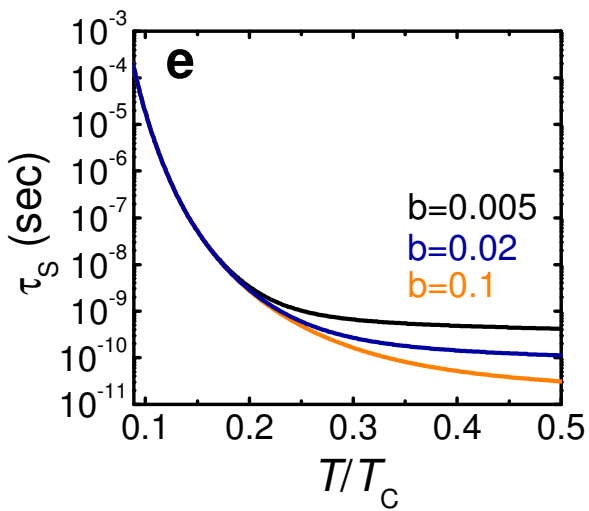

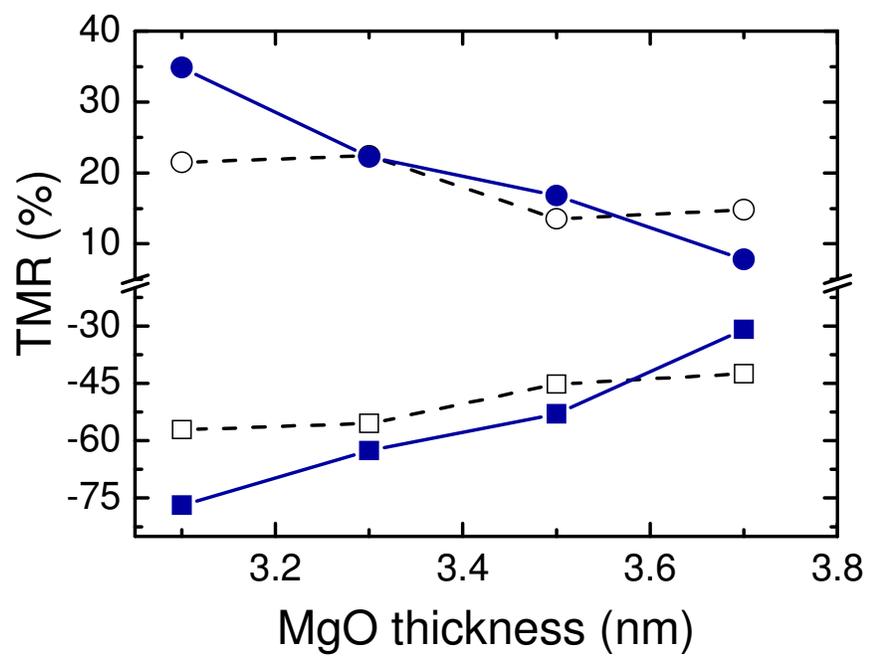

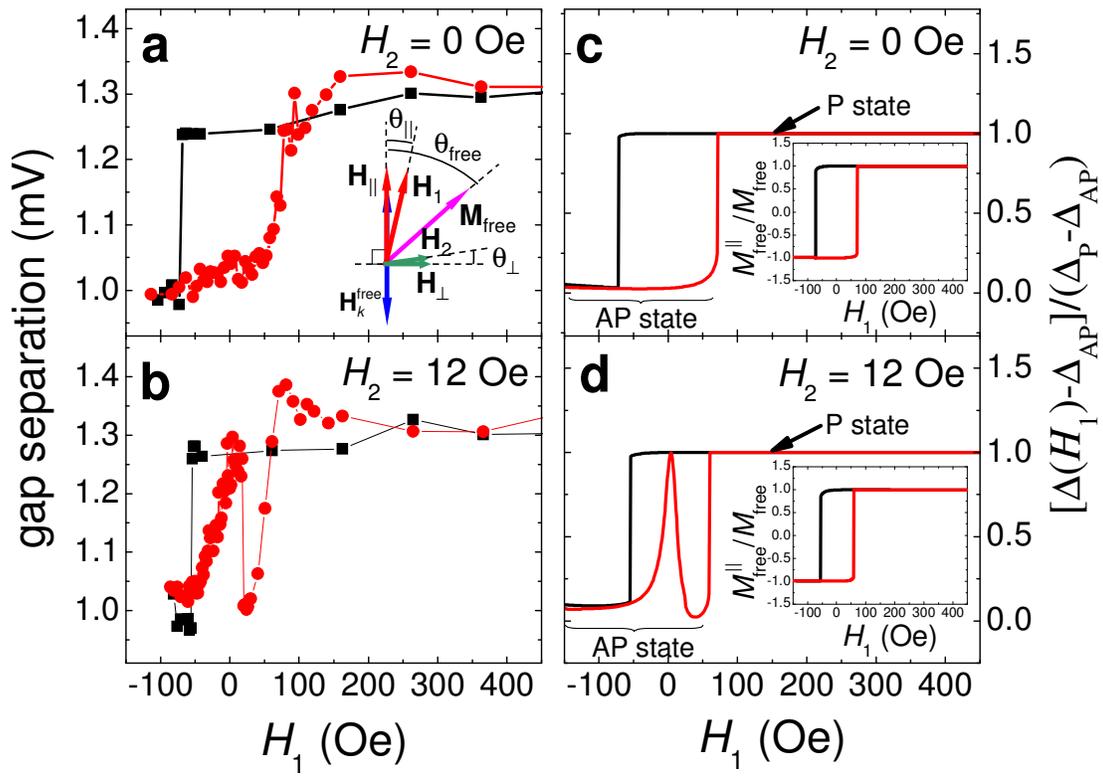